\documentclass[letterpaper, 10 pt, journal]{IEEEtran} % Comment this line out if you need a4paper

\IEEEoverridecommandlockouts                              % This command is only needed if
\usepackage{amsmath,mathrsfs}
\usepackage{amstext}
\usepackage{color,colortbl}
\usepackage[table]{xcolor}
\usepackage{amsfonts}
\usepackage{amssymb,amsbsy}
\usepackage{graphicx}
\usepackage{epstopdf}
\usepackage{url}
\usepackage[us]{datetime}
\usepackage{multirow}
\usepackage{cite}
\usepackage{comment}
\usepackage[thmmarks, amsmath]{ntheorem}
\usepackage[shortcuts,acronym]{glossaries}
\definecolor{Gray}{gray}{0.9}
\definecolor{LightCyan}{rgb}{0.88,1,1}
\definecolor{LightMagenta}{rgb}{1,0.88,1}
\definecolor{LightOrange}{rgb}{1,0.99,0.88}
\definecolor{LightGreen}{rgb}{0.9,1,0.8}
\definecolor{DarkOrange}{rgb}{0.98,0.91,0.71}
\definecolor{DarkGreen}{rgb}{0.67,0.88,0.69}

\newcommand{\sign}[1]{\textrm{sign}(#1)}

\allowdisplaybreaks[1]

\theorembodyfont{\normalfont}
\theoremseparator{:}
\newtheorem{claim}{Claim}%[section]
\newtheorem{remark}{Remark}%[section]
\newtheorem{proposition}{Proposition}%[section]
\theoremstyle{nonumberplain}
\theoremsymbol{$\square$}

\title{%\Large \bf
A Nonlinear Lateral Controller Design for Vehicle Path-following with an Arbitrary Sensor Location
}
\IEEEpubid{{\color{red} This paper is currently under review. This preprint is dedicated for REVIEW ONLY}.}

\author{Wubing~B.~Qin and Zhaojian Li% <-this % stops a space
\thanks{Manuscript revised \currenttime, \today.%
}
\thanks{W.~B.~Qin is with the Department of Mechanical Engineering, University of Michigan, Ann Arbor, MI 48109, USA. (Email: wubing@umich.edu).}%
\thanks{Zhaojian Li is with the Department of
Mechanical Engineering, Michigan State University, Lansing, MI, 48824,
USA. (Email: {lizhaoj1}@msu.edu).}
}
\begin{document}
%\date{}
\maketitle
\begin{abstract}
This paper investigates the lateral control problem in vehicular path-following when the feedback sensor(s) are mounted at an arbitrary location in the longitudinal symmetric axis. We point out that some existing literature has abused the kinematic bicycle model describing the motion of rear axle center for other locations, which may lead to poor performance in practical implementations. A new nonlinear controller with low-complexity and high-maneuverability is then designed that takes into account senor mounting location, driving comfort and transient response with large initial errors. Design insights and intuitions are also provided in detail. Furthermore, analysis on stability and tracking performance for the closed-loop system are studied, and conditions and guidelines are provided on the selection of control parameters. Comprehensive simulations are performed to demonstrate the efficacy of the proposed nonlinear controller for arbitrary sensor locations. Meanwhile, we also show that designing controllers ignoring the sensor location may lead to unexpected vehicular sway motion in non-straight paths.
\end{abstract}


\section{Introduction}

The problem of steering controller design for vehicular path-following  dates back to the 1950s \cite{Segel_1956, Fenton_1976, Shladover_JDSMC_1978, Ackermann_TCST_1995}, and many different control techniques \cite{Snider_2009, Samuel_IJCA_2016, Amer_JIRS_2016} have been developed thereafter. These techniques can be categorized into geometry-based approaches, control-based approaches, and learning-based approaches \cite{Bae2020, Ave2021}. Geometry-based approaches include pure pursuit \cite{Amidi_1990, Campbell_2007, Bidabad_NA_2009, Park_ETRI_2015,Andersen_AIM_2016, Elban_JVC_2018}, CF-pursuit \cite{Shan_ARS_2015}, vector pursuit \cite{Wit_2000}, and Stanley method \cite{Hoffmann_ACC_2007, AbdE_IJARS_2020}. In contrast,  control-based approaches utilize analysis and design from classical control theory \cite{Chatzikomis_2009, Marino_CEP_2011}, robust control \cite{Cuvenc_TCST_2004}, nonlinear control \cite{Sotelo_RAS_2003, Rossetter_2006, Talvala_Gerdes_2011, Laghrouche_2013, Shin_CEP_2015,Choi_Hedrick_2015_ECC}, optimal control \cite{Borrelli_IJVAS_2005, Falcone_TCST_2007, Falcone_IJRNC_2008, Raffo_2009, Brown_CEP_2017, XuPenTan_2021}, and fuzzy control \cite{Perez_IVS_2012, Rastelli_ASC_2015}.
These control strategies have been demonstrated to perform well in certain scenarios, such as when information about the rear axle center is obtainable, small steering/slip angles assumption holds, or computational cost is not problematic.

Nonetheless, with the increased level of autonomy \cite{SAE_J3016_2016}, recent efforts in automated vehicles have been mainly dedicated to handing all driving tasks. As a result, these advanced driving features typically overload the controllers with more than twenty tasks, including
adaptive cruise control (ACC),  forward collision warning/avoidance (FCW/FCA), lane centering (LC),  evasive steering assist (ESA), rear cross traffic alert (CTA), intersection collision avoidance (ICA), etc. The corresponding algorithms are implemented in middleware modules that bind the upper level planning/decision-making modules with the lower level actuator control modules. For cost reduction, the middleware is generally affordable and thus less powerful micro-controllers on production vehicles. Therefore, the automotive industry sets high expectations on those feature controllers, requiring high maneuverability, low cost, fast response, increased comfort, and enhanced safety.

However,  existing steering controllers rarely meet all these requirements simultaneously, and the issues are mainly three-fold. Firstly, for algorithms requiring online optimization such as model predictive control (MPC)-based approaches, it is a general practice to shorten the prediction horizon, to downgrade numerical precision, and to linearize models to fit the micro-controller platforms for real-time implementations. This undoubtedly leads to performance degradation. Secondly, for algorithms based on vehicle models with integrated tire models, poor performance might be observed on tight-curves at high speed due to model uncertainties. On the one hand, the design based on linearized models assumes small steering angle and slip angles, which results in non-negligible errors and unexpected behaviors for large steering efforts. On the other hand, the design based on nonlinear tire models utilizing brush model assumption or Pacejka magic formula \cite{Pacejka_1993, Rajamani_VDC, Popp_GVD, Ulsoy_Automotive_2012} requires great efforts in system/parameter identification. The uncertainties and errors in the underlying model assumption or parameter identification deteriorate the performance in those non-trivial scenarios. Thirdly, the strict requirements on vehicle states at the instant when steering controllers are initially engaged make them less capable in the presence of large errors, since transient responses to approach the desired path are seldom considered. Most algorithms only emphasize stability of the steady state solution that corresponds to close tracking of desired paths while leaving the task of ensuring transient responses to planning algorithms. This lack of consideration on transient responses in controller design also poses a threat to safety in the presence of disturbances/uncertainties or in switching scenarios since the vehicle states may temporarily be far from the steady state solution.

\IEEEpubidadjcol

To address the aforementioned issues and meet the expectations of automotive industry, we proposed a low-complexity nonlinear controller in \cite{Wubing_AMR_2021} that enables vehicles to execute a large variety of maneuvers for the path-following problem and takes into consideration the driving comfort and transient response with initial large errors. This nonlinear controller is applied to the single track models first developed through the Appellian approach \cite{Haim_1999}, and extended to scenarios integrating steering dynamics or incorporating longitudinal controllers. Meanwhile, we pointed out a caveat that some existing research has abused the kinematic bicycle model that only describes the motion of rear axle center. Also, it is mentioned that the proposed nonlinear controller is still only applicable to the scenario when sensory information is about the rear axle center.
Indeed, when the sensory information is about other locations, one can still utilize this type of controllers by transforming the information to that about rear axle center given vehicle states. However, this transformation amplifies measurement noises according to error propagation \cite{Ku_1966, Clifford_1973}, and may lead to unexpected behaviors due to large measurement noises.

To resolve this issue, in this paper we investigate more general scenarios  where the sensory information is about an arbitrary position along the vehicle symmetry axis, and design a new nonlinear controller that requires no transformation. The practical importance lies in the fact that the information needed for the path-following problem is typically obtained through GPS or cameras that are often mounted around rear-view mirrors or the center of mass. We show that the controller proposed in \cite{Wubing_AMR_2021} is a degenerated form in the aforementioned special case.
Another contribution is that we point out that the common design goals to maintain zero yaw angle error and zero lateral deviation are not simultaneously achievable on curvy roads for general sensor mounting positions. Instead, a curvature-dependent yaw angle error is needed to achieve zero lateral deviation and best tracking performance in such scenarios. Moreover, we demonstrate that vehicular sway motion will be generated on curvy roads if sensor positions are not considered, that is, controllers are designed based on the motion of rear axle center regardless of the actual sensor location.
This unexpected motion has been previously misunderstood as improper choices of control parameters. In comparison, the proposed controller is shown to provide satisfactory performance with proper settings according to our analysis on stability and tracking performance.

The remainder of this paper is organized as follows. In Section~\ref{sec:prob_statement}, we introduce the path-following problem in the case of an arbitrary sensor location along the longitudinal symmetry axis. A new nonlinear controller is then proposed in Section~\ref{sec:ctrl_design} with details and insights on the controller design, whereas analysis on stability and tracking performance are performed for the closed-loop system in Section~\ref{sec:analysis}. Comprehensive simulations are conducted in Section~\ref{sec:sim_res} to demonstrate the efficacy of the proposed controller. Finally, concluding remarks are drawn in Section~\ref{sec:conclusion}.

\section{Problem Statement \label{sec:prob_statement}}
Vehicle modeling and control have been extensively studied over the past decades and comprehensive results have been acquired. To obtain high-fidelity but low-complexity models that can provide insights on controller design, eight bicycle models are derived in \cite{Wubing_AMR_2021} which considers non-holonomic constraints in the Appellian framework. With these models, a nonlinear path-following controller is also presented to ensure that the rear axle center point can move along the given reference path, which is implemented in these models to demonstrate the effectiveness and extendability. However, it is noted that the proposed controller can only ensure satisfactory performance when the feedback information is about the rear axle center. This is rather restrictive as in practice the sensory information is provided by GPS devices or cameras, which are generally not mounted exactly at that specific location. Therefore, the goal of this paper is to generalize the controller design such that it is suitable for scenarios where sensors are mounted at an arbitrary location along the longitudinal symmetry axis. In the more general but less typical scenario where sensors are not mounted on the symmetry axis, one may need transformation to project the sensor information onto the closest point along the symmetry axis; otherwise constant lateral offset will be observed due to the lateral displacement of the sensor location. However, this non-typical scenario is beyond the scope of this paper.

\begin{figure}[!t]
\begin{center}
\includegraphics[scale=0.95]{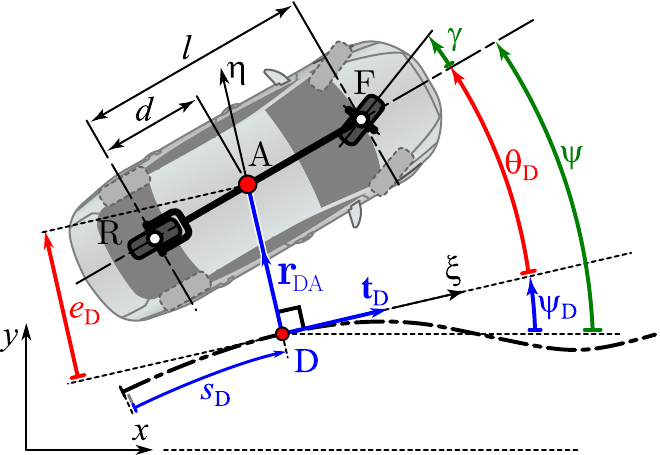}
\end{center}
\caption{Schematics of vehicle path-following problem. \label{fig:steer_ctrl}}
\end{figure}

In this paper, we consider the bicycle model \cite{Wubing_AMR_2021} with constrained longitudinal speed and assigned steering angle when the wheels are modeled as skates, that is,
\begin{equation}\label{eqn:EOM_xy}
\begin{split}
\dot{x}_{\rm A} &=V\Big(\cos\psi-\frac{d}{l}\sin\psi\tan\gamma\Big)\ , %\label{eqn:EOM_000_a}
\\
\dot{y}_{\rm A} &=V\Big(\sin\psi+\frac{d}{l}\cos\psi\tan\gamma\Big)\ , %\label{eqn:EOM_000_b}
\\
\dot{\psi} &= \frac{V}{l}\tan\gamma\ ,
%\label{eqn:EOM_000_c}
\end{split}
\end{equation}
where $V$ is the longitudinal speed, $l$ is the wheelbase length, $\gamma$ is the steering angle, $\psi$ is the vehicle yaw angle with respect to the $x$ axis, and A is the sensor mounting location whose distance to the center of rear axle is $d$; see Fig.~\ref{fig:steer_ctrl}. Here $x_{\rm A}$ and $y_{\rm A}$ are the coordinates of point A in the earth-fixed frame.
In this model, it is assumed that there is no side slip on the front and rear wheels, the steering angle is assigned, and the longitudinal speed is constant. We note an important property of the model that will be used later in this paper, that is,  setting $d=0$ results in the following lateral acceleration $a_{\rm R}^{\rm lat}$ at the rear axle center point R:
\begin{align}\label{eqn:lat_accel_R}
  a_{\rm R}^{\rm lat} &= -\ddot{x}_{\rm R} \sin\psi + \ddot{y}_{\rm R}  \cos\psi = \dfrac{V^{2}}{l}\tan\gamma\, ,
\end{align}
which only depends on the longitudinal speed and the steering angle. We remark that this property still holds when longitudinal speed is not a constant.

%%%%%%%%%%%%%%% begin figure %%%%%%%%%%%%%%%%%%%
%%%%%%%%%%%%%%% end figure %%%%%%%%%%%%%%%%%%%

As shown in Fig.~\ref{fig:steer_ctrl}, the vehicle aims to follow a given reference path depicted by the black dashed curve.
More specifically, the control design goal is to make point A follow this desired path. We use the positions (${{x}_{\rm A} , {y}_{\rm A}  }$) of the point A and the yaw angle $\psi$ as states to localize the vehicle in the ${(x,y)}$ plane. Point D marks the closest point to A along the given path. Moreover, $\mathbf{r}_{\rm DA}$ denotes the vector pointing from D to A and $\mathbf{t}_{\rm D}$ is the unit tangential vector of the path at point D. Observe that ${\mathbf{t}_{\rm D} \perp \mathbf{r}_{\rm DA}}$. The angle $\psi_{\rm D}$ indicates the direction of the tangential vector $\mathbf{t}_{\rm D}$ while the curvature of the path at point D is denoted by $\kappa_{\rm D}$. The path-reference frame $(\xi,\eta)$ travels along the path as the vehicle moves forward, and thus, the angle $\psi_{\rm D}(s_{\rm D})$ and the curvature $\kappa_{\rm D}(s_{\rm D})$ depends on the arc length $s_{\rm D}$. We assume that the information about the given path is known, namely, the tuple ${(x_{\rm D},\, y_{\rm D},\, \psi_{\rm D},\, \kappa_{\rm D})}$ are given as functions of the arc length $s_{\rm D}$.

To follow the desired path, a controller has to correct the lateral deviation and the yaw angle error with respect to the path. The lateral deviation can be calculated as
\begin{align}
e_{\rm D}&=(\mathbf{t}_{\rm D} \times \mathbf{r}_{\rm DA})\cdot \mathbf{k}\,,
\end{align}
which is positive when point D is on the left hand side of the path. Here, we use $\mathbf{k}$ to denote the unit vector along $z$ axis.
Then, by expressing the tangential vector $\mathbf{t}_{\rm D}$ with the angle $\psi_{\rm D}$, we can obtain the lateral deviation as
\begin{align}\label{eqn:lat_err}
  e_{\rm D}
    &=-({x}_{\rm A} -x_{\rm D})\sin\psi_{\rm  D}+({y}_{\rm A}  -y_{\rm D})\,\cos\psi_{\rm  D}\,.
\end{align}
Similarly, one can define the yaw angle error as
\begin{align}\label{eqn:heading_err}
  \theta_{\rm D}&=\psi-\psi_{\rm D}\,.
\end{align}
We remark that to ensure ${\theta_{\rm D}\in [-\pi,\,\pi)}$, one can generalize this definition as ${\theta_{\rm D}=\psi-\psi_{\rm D}-2\pi\left[\frac{\psi-\psi_{\rm D}}{2\pi}\right]}$, where $[\cdot]$ refers to the round function that rounds to the nearest integer. We will use this generalized definition in the simulations in Section~\ref{sec:sim_res}.

In order to determine the evolution of the lateral deviation and yaw angle error, one should first apply the nonlinear transformation to transform the coordinates from the $(x,y)$ earth-fixed frame to the $(\xi,\eta)$ path-reference frame. This transformation is detailed in \cite{Wubing_AMR_2021}, and \eqref{eqn:EOM_xy} is transformed into
\begin{align}\label{eqn:EOM_gen_point}
  \dot{s}_{\rm D}&=\dfrac{V}{1-e_{\rm D}\kappa_{\rm D}}\Big(\cos\theta_{\rm D}-\dfrac{d}{l}\tan\gamma\,\sin\theta_{\rm D}\Big)\,,\nonumber\\
  \dot{e}_{\rm D}&=V\Big(\sin\theta_{\rm D}+\dfrac{d}{l}\tan\gamma\,\cos\theta_{\rm D}\Big)\,,\\
  \dot{\theta}_{\rm D}&=\dfrac{V}{l}\tan\gamma-\dfrac{V\kappa_{\rm D}}{1-e_{\rm D}\kappa_{\rm D}}\Big(\cos\theta_{\rm D}-\dfrac{d}{l}\tan\gamma\,\sin\theta_{\rm D}\Big)\,, \nonumber
\end{align}
where the first equation characterizes the longitudinal motion of point D along the path, while the last two equations provide the evolution of lateral deviation and the yaw angle error with respect to the path.
Note again that the curvature $\kappa_{\rm D}(s_{\rm D})$ depends on the path coordinate $s_{\rm D}$, making the differential equations in \eqref{eqn:EOM_gen_point} all coupled.
After solving \eqref{eqn:EOM_gen_point}, the state $(s_{\rm D}, e_{\rm D}, \theta_{\rm D})$ can be transformed back to $({x}_{\rm A}, {y}_{\rm A}, \psi)$ by
\begin{equation}\label{eqn:xy_transf}
  \begin{split}
    {x}_{\rm A}  & =x_{\rm D} -e_{\rm D}\sin\psi_{\rm D}\,,
    \\
    {y}_{\rm A}   & =y_{\rm D} +e_{\rm D}\cos\psi_{\rm D}\,,
    \\
    \psi & = \psi_{\rm D} +\theta_{\rm D}\,.
  \end{split}
\end{equation}
Note that when point D coincides with the rear axle center R, i.e., $d=0$, the problem degenerates to the special case presented in \cite{Wubing_AMR_2021}.

\section{Controller Design \label{sec:ctrl_design}}

In this section, we focus on the controller design for the path-following problem defined in Section~\ref{sec:prob_statement}. Specifically, in Section \ref{sec:ctrl_design_ultimate} we first present the controller design for an arbitrary sensor mounting location along the vehicle symmetry axis. Then in Section~\ref{sec:design_details} we provide further insights and details on the design of the developed feedforward control and feedback control. Readers, who are not interested in these details, may jump to Section~\ref{sec:analysis} for performance analysis or to Section~\ref{sec:sim_res} for simulation results after reading Section~\ref{sec:ctrl_design_ultimate}.

\subsection{Nonlinear Controller Design\label{sec:ctrl_design_ultimate}}
Given the dynamic system \eqref{eqn:EOM_xy} or \eqref{eqn:EOM_gen_point}, we assume the steering angle $\gamma$ can track any desired value $\gamma_{\rm des}$, that is,
\begin{align}\label{eqn:steer_angle_track}
  \gamma &= \gamma_{\rm des}.
\end{align}
The goal of the path-following controller design is to determine the desired steering angle $\gamma_{\rm des}$ such that point A on the vehicle can follow a given path; see Fig.~\ref{fig:steer_ctrl}.

We propose the path-following controller
\begin{align}\label{eqn:lateral_controller}
  \gamma_{\rm des} & = \gamma_{\rm ff} + \gamma_{\rm fb}\,,
\end{align}
which consists of a feedforward control law
\begin{align}
  \gamma_{\rm ff} &= \arctan \dfrac{l\,\kappa_{\rm D}}{\sqrt{1-(d\,\kappa_{\rm D})^{2}}}\,, \label{eqn:steer_controller_ff_sum}
\end{align}
and a feedback control law
\begin{align}\label{eqn:steer_controller_fb_sum}
  \gamma_{\rm fb} &= g \Big( k_{1}\big(\theta_{\rm D}-\theta_{0}+\arctan (k_{2}\,e_{\rm D})\big)\Big)\,.
\end{align}
The feedforward control essentially provides the estimated steering angle necessary to handle a given road curvature $\kappa_{\rm D}$ whereas the feedback control makes corrections based on lateral deviation $e_{\rm D}$ and yaw angle error $\theta_{\rm D}-\theta_{0}$. In \eqref{eqn:steer_controller_fb_sum},
\begin{align}\label{eqn:theta0_des}
  \theta_{0} & = -\arcsin(d\,\kappa_{\rm D})\,
\end{align}
is the desired yaw angle error, ($k_{1}$, $k_{2}$) are the tunable control gains, and $g (x)$ denotes a wrapper function with the following properties:
\begin{enumerate}
  \item It is continuously differentiable and monotonically increasing over $\mathbb{R}$.
  \item It is an odd function, i.e., ${g(x)=-g(-x)}$ for ${x\in \mathbb{R}}$.
  \item It is  bounded,  i.e., there exists a constant $g_{\rm sat}$ such that
  ${|g(x)| \le g_{\rm sat}}$ for ${x\in \mathbb{R}}$.
  \item Its derivative decreases monotonically for ${x\in \mathbb{R}_{\geq0}}$ such that ${g'(0)=1}$ and ${\lim\limits_{x \to +\infty}  g'(x) = 0}$.
\end{enumerate}
In this paper we use the wrapper function
\begin{equation}
    g(x) =\dfrac{2\,g_{\rm sat}}{\pi}\arctan \Big(\dfrac{\pi}{2\,g_{\rm sat}} x\Big)\,, \label{eqn:satfunction}
\end{equation}
where the bound $g_{\rm sat}$ is set to the the maximum allowable steering angle
\begin{align}
  \gamma_{\rm sat} = \min\left\{\gamma_{\max},\, \arctan\Big(\dfrac{a_{\max}^{\rm lat}\,l}{V^2}\Big)\right\}\,.
  \label{eqn:max_allow_steer_ang}
\end{align}
This choice of bound ensures the resulting steering angle and lateral acceleration will not exceed the physical steering angle limit $\gamma_{\max}$ and the maximum allowed lateral acceleration $a_{\max}^{\rm lat}$, respectively.

\subsection{Design insights and further details\label{sec:design_details}}

%%%%%%%%%%%%%%% begin figure %%%%%%%%%%%%%%%%%%%
\begin{figure}[!t]
  \centering
  \includegraphics[scale=0.6]{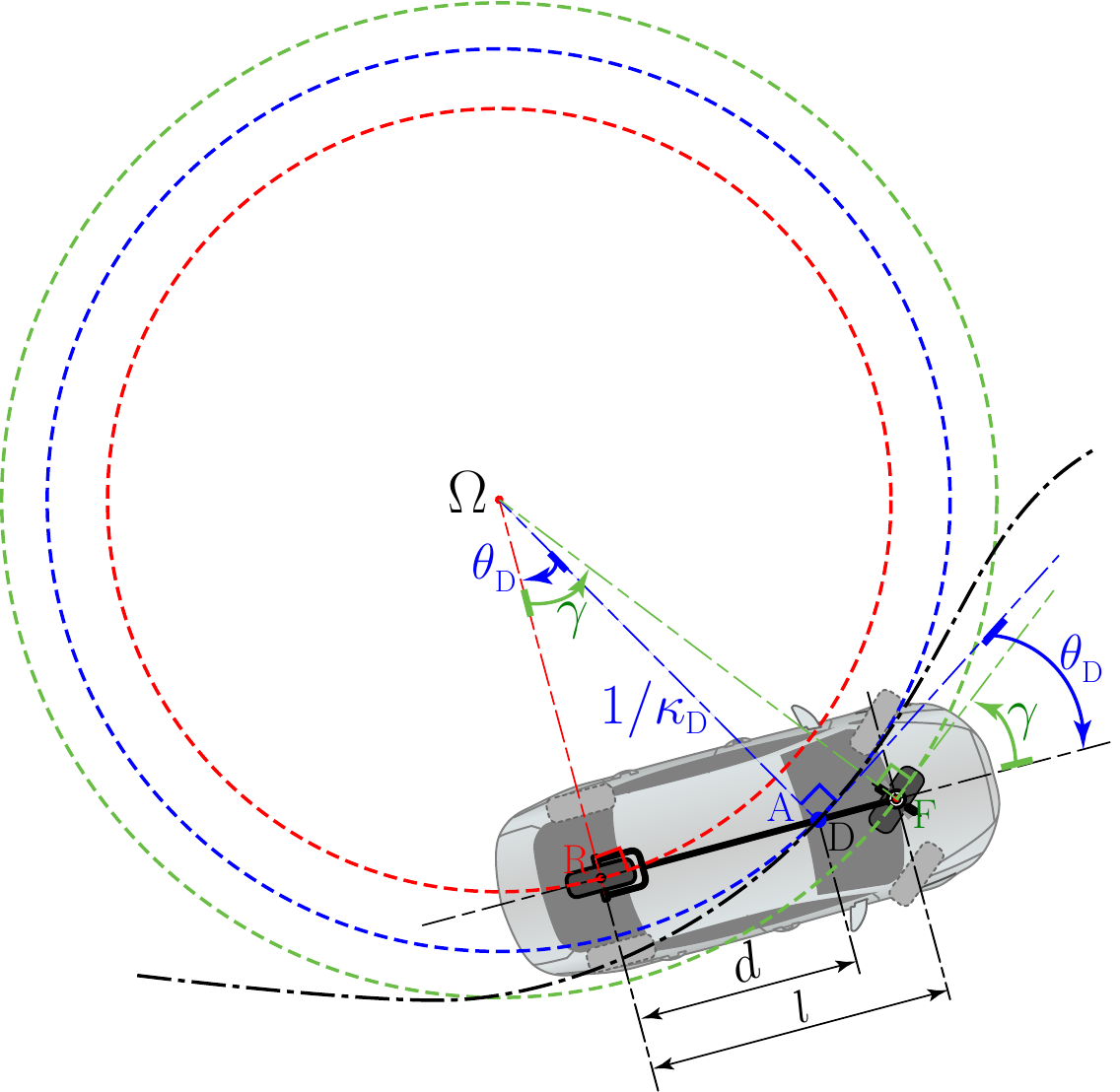}
  \caption{Vehicle point A follows a given path ideally.
  \label{fig:circ_path_steer}}
\end{figure}
%%%%%%%%%%%%%%% end figure %%%%%%%%%%%%%%%%%%%

In this part, we provide insights and details on the controller presented above. In particular, Fig.~\ref{fig:circ_path_steer} depicts the scenario where point A is moving along the desired path ideally (i.e., A always coincides with D). The black dashed curve represents the given path, while the red, green and blue circles represent the osculating circles at the points of rear wheel, front wheel and point A, respectively. $\Omega$ is the instantaneous center of rotation. Thus, from geometry and kinematics, we obtain the following facts:
\begin{enumerate}
  \item the osculating circles of point R, F and A are concentric, and centered at $\Omega$.
  \item the radius of the osculating circle at A must be equal to the radius of curvature at point D to ensure perfect tracking, i.e., $|\Omega A| = 1/|\kappa_{\rm D}|$.
  \item the velocity vector of rear/front wheel is parallel to the tangential direction at point R/F with respect to its osculating circle because of the no-side-slip condition on the rear/front wheel.
  \item there exists a curvature-dependent yaw angle error $\theta_{0}$ (cf.~\eqref{eqn:theta0_des}) between the vehicle heading direction and the tangential direction at point D with respect to its osculating circle.
\end{enumerate}
Based on these facts, next we will provide more insights on the feedforward and feedback controller design.
\subsubsection{Feedforward Control Design\label{sec:feedforward}}

Based on facts (1, 2, 3), the required steering angle to ensure the perfect tracking can be solved in terms of path information (road curvature) and vehicle information (wheelbase length), which can be used as the feedforward term. That is,
\begin{align}
  \tan\gamma_{\rm ff} &=\sign{\kappa_{\rm D}}\dfrac{l}{\sqrt{\kappa_{\rm D}^{-2}-d^{2}}}\,,&
\end{align}
yielding the feedforward term \eqref{eqn:steer_controller_ff_sum}. We remark that \eqref{eqn:steer_controller_ff_sum} degenerates to the feedforward term
\begin{align}
  \hat{\gamma}_{\rm ff} &= \arctan (l\,\kappa_{\rm D})\,, \label{eqn:steer_controller_ff_D_0}
\end{align}
in \cite{Wubing_AMR_2021} when the sensors are mounted at the rear axle center, i.e., $d=0$. Note that to determine the feedforward law, a geometrical approach is applied here instead of the analytical approach presented in \cite{Wubing_AMR_2021}, since the latter one is less straightforward when more complex dynamics are involved in \eqref{eqn:EOM_gen_point}.

\begin{figure}[!t]
  \centering
  \includegraphics[scale=1]{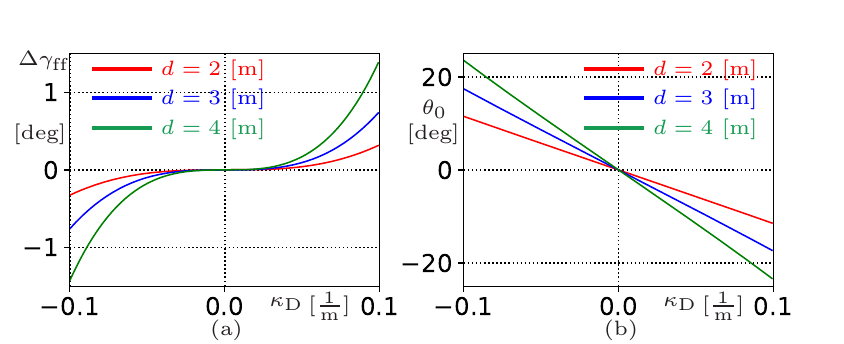}
  \caption{(a) Feedforward term error $\Delta\gamma_{\rm ff}$ in \eqref{eqn:ff_D_diff}. (b) Desired yaw angle error $\theta_{0}$ in \eqref{eqn:theta0_des}. \label{fig:terms_correct_D}}
\end{figure}

Also, we highlight that when sensors are not mounted at the rear axle center, i.e., $d\ne 0$, neglecting the effect of mounting position leads to an error in the feedforward term
\begin{align}\label{eqn:ff_D_diff}
  \Delta\gamma_{\rm ff} & =\arctan \dfrac{l\,\kappa_{\rm D}}{\sqrt{1-(d\,\kappa_{\rm D})^{2}}} -\arctan (l\,\kappa_{\rm D})\,,
\end{align}
which will cause unexpected vehicular sway motion when road curvature varies. This will be demonstrated in Section~\ref{sec:sim_res}.
Fig.\ref{fig:terms_correct_D}(a) shows the error \eqref{eqn:ff_D_diff} when $d=2$, $3$ and $4$ [m] and $l=2.57$ [m]. Indeed, this error is tiny when path curvature is small, implying that (\ref{eqn:steer_controller_ff_D_0}) might still be able to handle non-tight curves. However, (\ref{eqn:steer_controller_ff_D_0}) becomes less capable as road curvature increases due to the increase of $\Delta\gamma_{\rm ff}$.

%%%%%%%%%%%%%%% begin figure %%%%%%%%%%%%%%%%%%%
%%%%%%%%%%%%%%% end figure %%%%%%%%%%%%%%%%%%%

The feedforward term \eqref{eqn:steer_controller_ff_sum} can only predict the desired steering angle perfectly under the conditions that the measurements are ideally clean, the model is ideally accurate, the initial state is precisely on the path, and there exists no other disturbance. However, those requirements cannot be satisfied in practice. Therefore, we need to design a feedback controller to ensure good tracking performance despite model uncertainties and measurement imperfections..

%%%%%%%%%%%%%%%%%%%%%%%%%%%%%%%%%%%%%%%%%%%%%%%%%%%%%%%%%%%%%%%%%%%%%%
\subsubsection{Feedback Control Design}
%%%%%%%%%%%%%%%%%%%%%%%%%%%%%%%%%%%%%%%%%%%%%%%%%%%%%%%%%%%%%%%%%%%%%%
In feedback control design, the first step is to acquire the desired states from the dynamics.
In \eqref{eqn:EOM_gen_point}, the lateral dynamics are characterized by the evolution of lateral deviation $e_{\rm D}$ and yaw angle error $\theta_{\rm D}$. Based on the ideal scenario depicted in Fig.~\ref{fig:circ_path_steer} and fact (4), it is clear that the desired lateral deviation is zero, and the desired yaw angle error is $\theta_{0}$. We highlight that this is different from most lateral controllers in the literature where the desired lateral deviation and yaw angle error are both zero. As a matter of fact, one can easily observe from Fig.~\ref{fig:circ_path_steer} that the lateral deviation and yaw angle error cannot be zero simultaneously in the ideal scenario. Also, it can be verified that in \eqref{eqn:EOM_gen_point} when $\kappa_{\rm D}\ne 0$, there exists no $\gamma$ such that $e_{\rm D}\equiv 0$ and $\theta_{\rm D}\equiv 0$ hold simultaneously. This implies that one cannot design a controller on $\gamma$ for curvy roads such that a vehicle can maintain the state corresponding to zero lateral deviation and zero yaw angle error.

In fact, the ultimate goal of path-following controllers is to make the vehicle follow the given path without lateral deviation, and the yaw angle error $\theta_{0}$ is  needed to achieve this goal on curvy roads.
We remark that  when the sensors are mounted at the rear axle center, i.e., $d=0$, this desired yaw angle error $\theta_{0}$ becomes zero, and the feedback term \eqref{eqn:steer_controller_fb_sum} degenerates to the one presented in \cite{Wubing_AMR_2021}, that is,
\begin{align}
  \hat{\gamma}_{\rm fb} &= g \Big( k_{1}\big(\theta_{\rm D}+\arctan (k_{2}\,e_{\rm D})\big)\Big)\,. \label{eqn:steer_controller_fb_D_0}
\end{align}

Fig.~\ref{fig:terms_correct_D}(b) plots this desired yaw angle error $\theta_{0}$ against curvatures when $d=2$, $3$ and $4$ [m], respectively. It can be seen that this term is crucial since it varies significantly when curvature $\kappa_{\rm D}$ or location $d$ changes. Ignoring this term will lead to undesired behaviors, which will be demonstrated in the simulations in Section \ref{sec:sim_res}.

To gain more insights, in the following we study the behaviors of controller \eqref{eqn:steer_controller_fb_sum} in the case of straight-path following.
As explained above, the desired yaw angle error $\theta_{0}$ is specifically designed for curvy roads to achieve zero lateral deviation.
When the road is straight, the desired yaw angle error $\theta_{0}$ becomes zero, and the feedback term \eqref{eqn:steer_controller_fb_sum} degenerates to \eqref{eqn:steer_controller_fb_D_0} as well.
This nonlinear feedback control law allows the vehicle to correct the steering angle for both small and large values of the errors $e_{\rm D}$ and $\theta_{\rm D}$. For small errors one may neglect the nonlinearities and obtain the linearized controller
\begin{align}
  \gamma^{0}_{\rm fb} & = k_1  \theta_{\rm D} + k_1 k_2 e_{\rm D}\, , \label{eqn:steer_linear_fb}
\end{align}
which is widely used in the literature (see e.g., \cite{Chatzikomis_2009}). However,
as demonstrated below, for large errors this linear feedback controller may produce unwanted behaviors.

\begin{figure}[!t]
  \centering
  \includegraphics[scale=0.9]{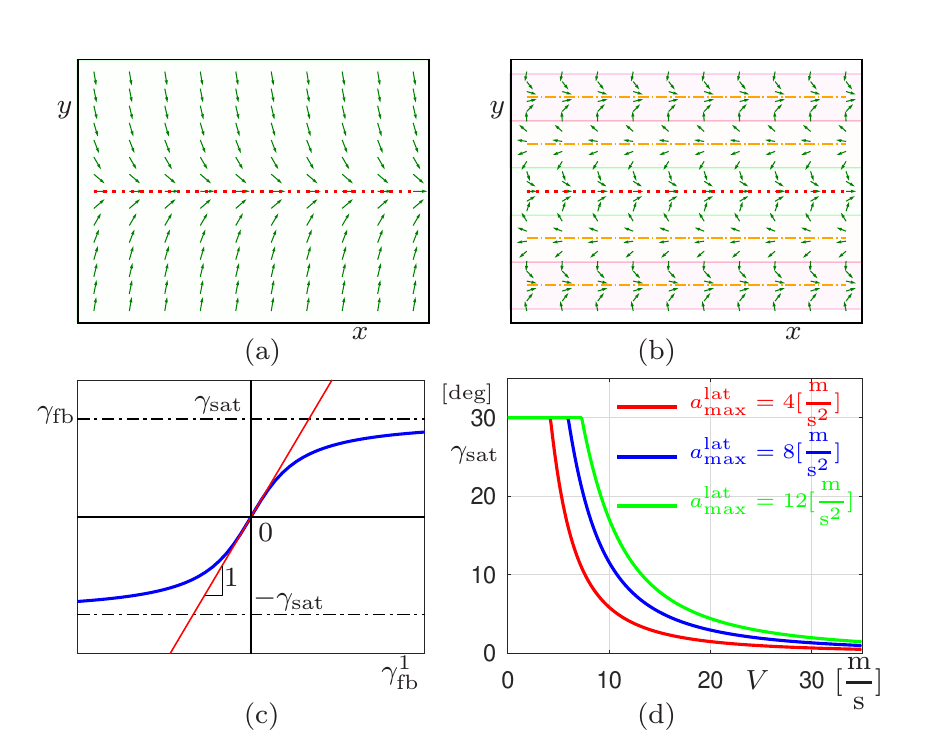}
  \caption{(a) Desired vehicle heading with nonlinear controller \eqref{eqn:steer_nonlinear_fb}. (b) Desired vehicle heading with linear controller  \eqref{eqn:steer_linear_fb}. (c) Wrapper function  \eqref{eqn:satfunction}. (d) Maximum allowable steering angle \eqref{eqn:max_allow_steer_ang} at different speed.
  \label{fig:fun_arctan_steer_max}}
\end{figure}

If omitting the wrapper function in \eqref{eqn:steer_controller_fb_sum},
we obtain
\begin{align}
  \gamma^{1}_{\rm fb}  =k_{1}\big(\theta_{\rm D}+\arctan (k_{2}\,e_{\rm D})\big)\, . \label{eqn:steer_nonlinear_fb}
\end{align}
 One may interpret this controller as
\begin{align}
  \gamma^{1}_{\rm fb}  &=k_{1}\big(\theta_{\rm D}-\theta^{\rm des}\big)\, , \\
  \theta^{\rm des}  &= - \arctan (k_{2}\,e_{\rm D})\, ,\label{eqn:desired_heading}
\end{align}
implying that it attempts to achieve the desired yaw angle error $\theta^{\rm des}$ based on the lateral deviation $e_{\rm D}$. We will validate this interpretation with simulations in Section~\ref{sec:sim_res}. Fig.~\ref{fig:fun_arctan_steer_max}(a) depicts the desired heading based on $\theta^{\rm des}$ as a field of green arrows in the $(x,y)$ plane where the desired path is given by the red dashed line. Notice that when the vehicle is far from the path, the desired heading points toward the path since
$ e_{\rm D}\rightarrow \pm \infty$ yields $\theta^{\rm des} \rightarrow \mp \pi/2$.

Similarly, one may interpret the linear controller \eqref{eqn:steer_linear_fb} as an effort to achieve the desired yaw angle error
\begin{align}
  \theta^{\rm des,0}  = - k_{2}\,e_{\rm D}\, ,
\end{align}
depending on the lateral deviation $e_{\rm D}$. The field of green arrows depicted in Fig.~\ref{fig:fun_arctan_steer_max}(b) show that this may lead to wrong desired headings when the vehicle is far from the path due to the $2\pi$ periodicity of the angle. For example, the linear controller forces the vehicle to move in parallel with the path when $e_{\rm D} = n \pi/k_2$, $n=\pm 1, \pm 2, \ldots$.

%%%%%%%%%%%%%%% begin figure %%%%%%%%%%%%%%%%%%%
%%%%%%%%%%%%%%% end figure %%%%%%%%%%%%%%%%%%%

One may notice that the controller \eqref{eqn:steer_nonlinear_fb} uses the same gain for small and large errors. However, in practice larger gains are preferred for small errors to achieve better tracking performance and smaller gains are preferred for large errors to avoid ``overreaction" and potential oscillations. To resolve this conflict, gain scheduling are typically utilized such that different gains can be applied for errors in different ranges. However, this technique requires strenuous tuning and stitching to ensure satisfactory performance and smooth transition across ranges. Alternatively, we propose a new method that can effectively decrease the gain when necessary. In particular, the wrapper function $g(x)$ is designed for this purpose. As indicated in Fig.~\ref{fig:fun_arctan_steer_max}(c), the derivative of the wrapper function $g(x)$ monotonically decreases with respect to $|x|$. As a result, the controller \eqref{eqn:steer_controller_fb_sum} utilizing this wrapper function will decrease the gain accordingly based on the error magnitude.

Regarding the allowable steering angle $\gamma_{\rm sat}$ of the feedback controller \eqref{eqn:steer_controller_fb_sum}, one can simply choose a value that is smaller than the physical steering angle limit $\gamma_{\max}$ of the vehicle. However, when the vehicle speed is high, this may lead to passenger discomfort and even roll-over hazards, which are related to resulting lateral acceleration. To avoid this, we include the lateral acceleration in the determination of the allowable steering angle. Indeed, different locations on the vehicle possess different lateral accelerations, but those at locations other than rear axle center is more complicated and the value at the rear axle center provides a baseline. Thus, we pick the rear axle center point.
Considering the scenario when the vehicle is approaching the given path with large initial errors, the desired steering angle is dominated by the feedback term, i.e., $\gamma_{\rm des} \approx \gamma_{\rm fb}$. The lateral acceleration can be approximated as
\begin{align}
  a_{\rm R}^{\rm lat} \approx \frac{V^{2}}{l}\tan \gamma_{\rm fb} < \frac{V^{2}}{l}\tan \gamma_{\rm sat}\,,
\end{align}
cf.~(\ref{eqn:lat_accel_R}, \ref{eqn:steer_angle_track}).
Thus, to ensure that the lateral acceleration caused by steering cannot exceed maximum allowed value $a_{\max}^{\rm lat}$ , and the steering angle cannot exceed the physical steering angle limit $\gamma_{\max}$, we set the maximum allowable steering angle as \eqref{eqn:max_allow_steer_ang}.
Fig.~\ref{fig:fun_arctan_steer_max}(d) shows this angle as a function of the longitudinal speed for different lateral acceleration limits $a_{\max}^{\rm lat}$ when $l=2.57$ [m] and $\gamma_{\max}=30$ [deg].

\section{Analysis \label{sec:analysis}}
In this section, we study the stability of the closed-loop system (\ref{eqn:EOM_gen_point}, \ref{eqn:steer_angle_track}) with the controller (\ref{eqn:lateral_controller}-\ref{eqn:theta0_des}) in the general case where sensors are mounted at an arbitrary location along the vehicle symmetry axis. Here, we assume that the vehicle is moving forward, i.e., $V>0$.

To study the dynamic tracking performance of the controller, we define the difference
\begin{align}\label{eqn:theta_hat}
  \hat{\theta}_{\rm D}&=\theta_{\rm D}-\theta_{0}\,,
\end{align}
as a state, and rewrite the dynamics \eqref{eqn:EOM_gen_point} as
\begin{align}\label{eqn:EOM_gen_point_transf}
  \dot{s}_{\rm D}&=\frac{V}{1-e_{\rm D}\kappa_{\rm D}}\Big(\cos(\hat{\theta}_{\rm D}+\theta_{0})-\tfrac{d}{l}\tan\gamma\,\sin(\hat{\theta}_{\rm D}+\theta_{0})\Big)\,,\nonumber\\
  \dot{e}_{\rm D}&=V\Big(\sin(\hat{\theta}_{\rm D}+\theta_{0})+\tfrac{d}{l}\tan\gamma\,\cos(\hat{\theta}_{\rm D}+\theta_{0})\Big)\,,\\
  \dot{\hat{\theta}}_{\rm D}&=\dfrac{V}{l}\tan\gamma +\dfrac{d\, \dot{\kappa}_{\rm D}}{\sqrt{1-d^{2}\kappa_{\rm D}^{2}}}\,\nonumber\\
  &-\dfrac{V\kappa_{\rm D}}{1-e_{\rm D}\kappa_{\rm D}}\Big(\cos(\hat{\theta}_{\rm D}+\theta_{0})-\tfrac{d}{l}\tan\gamma\,\sin(\hat{\theta}_{\rm D}+\theta_{0})\Big)\,. \nonumber
\end{align}

One can verify that the closed-loop system (\ref{eqn:EOM_gen_point_transf}, \ref{eqn:steer_angle_track}-\ref{eqn:theta0_des}) possesses the desired solution
\begin{align}\label{eqn:equil_D}
  s_{\rm D}^{\ast}  &=\dfrac{V t}{\sqrt{1-(d\,\kappa_{\rm D}^{\ast})^{2}}}\ ,&
  e_{\rm D}^{\ast}  &=0\ , &
  \hat{\theta}_{\rm D}^{\ast}  &=0\ ,
\end{align}
when $\kappa_{\rm D}^{\ast} := \kappa(s_{\rm D}^{\ast}) $ is constant, which corresponds to the desired motion when point A on the vehicle follows the path perfectly. We highlight the following facts:
\begin{enumerate}
  \item When sensors are mounted at the rear axle center (i.e., $d=0$), \eqref{eqn:equil_D} is always a solution to the closed-loop system regardless of the path. That is, the system can in principle follow any path without tracking errors.
  \item When sensors are not mounted at the rear axle center (i.e., $d\ne 0$), \eqref{eqn:equil_D} is only one particular solution to the closed-loop system when $\kappa_{\rm D}^{\ast}$ is constant. That is, the system is only able to follow paths with constant curvatures (straight/circular paths) without tracking errors.
\end{enumerate}
In the following we analyze the stability of  solution \eqref{eqn:equil_D} and characterize tracking errors when road curvature varies. Indeed, one can analyze the nonlinear system directly with Lyapunov method, which involves the challenging task of finding a Lyapunov function. Instead, we exploit an indirect method where we study its linearized system as the stability of the linearized system is topologically equivalent to the local stability of the corresponding nonlinear system. In addition, the linearized system can provide insights on the tracking performance when road curvature varies in the neighborhood of a nominal constant value.

We assume that the road curvature $\kappa_{\rm D}$ varies around a nominal value $\kappa_{\rm D}^{\ast}\equiv\kappa_{0}$, and view the perturbations
\begin{align}\label{eqn:perturb_kappa_D}
 \tilde{\kappa}_{\rm D} &= \kappa_{\rm D}-\kappa_{0}\, ,&
 \dot{\tilde{\kappa}}_{\rm D} &= \dot{\kappa}_{\rm D}\, ,
\end{align}
as the inputs to the closed-loop system.
By defining the perturbations of states around the nominal solution \eqref{eqn:equil_D} as
\begin{align}\label{eqn:perturb_D}
 \tilde{s}_{\rm D} = s_{\rm D}-s_{\rm D}^{*}\, ,\quad
 \tilde{e}_{\rm D} = e_{\rm D}-e_{\rm D}^{*}\, ,\quad
 \tilde{\theta}_{\rm D} = \hat{\theta}_{\rm D}-\hat{\theta}_{\rm D}^{*}\, ,
\end{align}
and linearizing the closed-loop system (\ref{eqn:EOM_gen_point_transf}, \ref{eqn:steer_angle_track}-\ref{eqn:theta0_des}), one can obtain
\begin{equation}\label{eqn:closed_lin_dyn_D}
\begin{split}
 \dot{\tilde{s}}_{\rm D}&= \dfrac{V \kappa_{0}}{\lambda_{1}}(1+\dfrac{\lambda_{2}}{\lambda_{1}}\dfrac{d^{2}}{l}k_{1}k_{2})\tilde{e}_{\rm D}
 +V \kappa_{0}\dfrac{\lambda_{2}}{\lambda_{1}^{2}}\dfrac{d^{2}}{l}k_{1} \tilde{\theta}_{\rm D}\\
 &+V \kappa_{0}\dfrac{d^{2}}{\lambda_{1}^{3}}\tilde{\kappa}_{\rm D}\, ,
 \\
 \dot{\tilde{e}}_{\rm D} &= \dfrac{V\, d}{l}\dfrac{\lambda_{2}}{\lambda_{1}}k_{1}k_{2}\tilde{e}_{\rm D}+\dfrac{V}{\lambda_{1}}(1+\dfrac{d}{l}\lambda_{2}k_{1})\tilde{\theta}_{\rm D}\,,\\
 \dot{\tilde{\theta}}_{\rm D} &= \dfrac{V}{l}(\lambda_{2} k_{1}k_{2}-\dfrac{l}{\lambda_{1}}\kappa_{0}^{2})\tilde{e}_{\rm D}+\frac{V\lambda_{2}}{l}k_{1}\tilde{\theta}_{\rm D}+\dfrac{d}{\lambda_{1}}\dot{\tilde{\kappa}}_{\rm D}\, ,\\
\end{split}
\end{equation}
where
\begin{align}\label{eqn:lambda_1_2}
  \lambda_{1} & = \sqrt{1-d^{2}\kappa_{0}^{2}}\,,&
  \lambda_{2} & = 1+(l^{2}-d^{2})\kappa_{0}^{2}\,.
\end{align}
In practice, the road is always constructed within the vehicle's steering capability, that is, there is no tight curves that vehicles cannot follow even by maximizing steering angles. Actually this is a very soft condition since in practice besides the consideration of vehicle capability, the road curvatures are deliberately designed to avoid large lateral accelerations and roll-over hazards at designated speed limit. This soft condition yields the following claim.
\begin{claim}\label{claim:lamda2_pos}
  Given the physical maximum of the steering angle $\gamma_{\max}$, the following always holds:
  \begin{enumerate}
    \item $|\kappa_{0}|\le \overline{\kappa}_{0}$ where $\overline{\kappa}_{0}=\dfrac{\tan\gamma_{\max}}{\sqrt{l^{2} +d^{2} \tan^{2}\gamma_{\max}}}$.
        \vspace{1pt}
    \item $\lambda_{1} \in [\,\underline{\lambda}_{1}, 1]$ where $\underline{\lambda}_{1} =\dfrac{l}{\sqrt{l^{2} +d^{2} \tan^{2}\gamma_{\max}}}>0$.
        \vspace{1pt}
    \item $\lambda_{2} >0 $.
  \end{enumerate}
  \IEEEproof See Appendix~\ref{append:proof_claim_1}.
\end{claim}

From \eqref{eqn:closed_lin_dyn_D}, one may observe that the last two equations are decoupled from the first one, and they form the reduced linear system for lateral control.
By defining the state and input as
\begin{align}
  \mathbf{x} & = \begin{bmatrix}
  \tilde{e}_{\rm D} \\ \tilde{\theta}_{\rm D}
  \end{bmatrix}\,, &
  \mathbf{u} & = \dot{\tilde{\kappa}}_{\rm D}\,.
\end{align}
and selecting the output as
\begin{align}
  \mathbf{y} = \tilde{e}_{\rm D}\,,
\end{align}
we obtain the standard linear state space model
\begin{equation}
\begin{split}\label{eqn:lin_model}
  \dot{\mathbf{x}} &= \mathbf{A}\mathbf{x}+\mathbf{B}\mathbf{u}\,,\\
  \mathbf{y} &= \mathbf{C}\mathbf{x}\,,
\end{split}
\end{equation}
where
\begin{equation}
\begin{split}
  \mathbf{A} &=
\begin{bmatrix}
  \frac{V\, d}{l}\frac{\lambda_{2}}{\lambda_{1}}k_{1}k_{2} & \frac{V}{\lambda_{1}}(1+\frac{d}{l}\lambda_{2}k_{1})
  \\
  \frac{V}{l}(\lambda_{2} k_{1}k_{2}-\frac{l}{\lambda_{1}}\kappa_{0}^{2}) & \frac{V\lambda_{2}}{l}k_{1}
  \end{bmatrix}\,,
  \\
  \mathbf{B}&=\begin{bmatrix}
  0\\ \frac{d}{\lambda_{1}}
  \end{bmatrix}\,, \qquad
  \mathbf{C} = \begin{bmatrix}
  1 &0
  \end{bmatrix}\,.
\end{split}
\end{equation}

\begin{figure}[!t]
  \centering
  \includegraphics[scale=1]{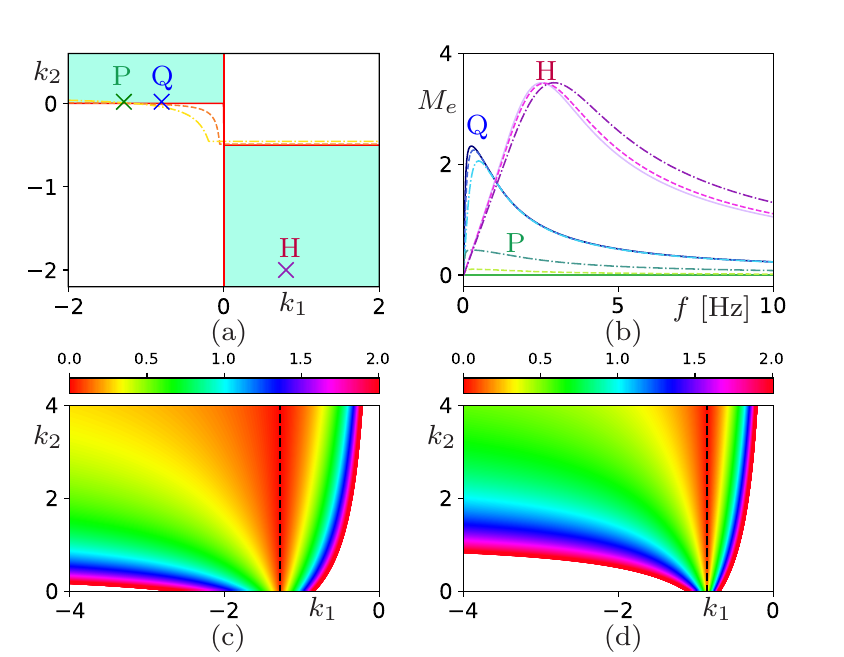}
  \caption{(a) Stability region in $(k_{1},\, k_{2})$-plane. (b) Amplification ratio $M(\omega)$ for points P, Q and H for different $\kappa_{0}$. (c) Contour plot of $M_{\max}$ in $(k_{1},\, k_{2})$-plane when $d=2$ [m]. (d) Contour plot of $M_{\max}$ in $(k_{1},\, k_{2})$-plane when $d=3$ [m].
  \label{fig:stab_diag}}
\end{figure}

The characteristic equation is
\begin{equation}
\begin{split}
  \det&(s\,\mathbf{I}-\mathbf{A}) =s^{2} -\tfrac{V k_{1}\lambda_{2}}{l\lambda_{1}} (\lambda_{1}+d\, k_{2})s
  -\tfrac{V^{2}\lambda_{3}}{l\lambda_{1}^{2}}=0\ ,
  \label{eqn:char_eqn}
\end{split}
\end{equation}
where
\begin{align}
\lambda_{3} = \lambda_{1}\lambda_{2} k_{1} k_{2}-d \kappa_{0}^{2} \lambda_{2} k_{1}-l \kappa_{0}^{2}\,,
\end{align}
and $s\in \mathbb{C}$ is the eigenvalue of the system. To ensure stability, all eigenvalues must lie in the left half of the complex plane.
Applying Routh-Hurwitz criterion, we can obtain the necessary and sufficient condition to ensure stability of the linearized system \eqref{eqn:lin_model}, that is,
\begin{align}\label{eqn:stab_cond_nec_suff}
  k_{1} (\lambda_{1}+d\, k_{2}) &<0\ ,&
  \lambda_{3}&<0\ .
\end{align}
Fig.~\ref{fig:stab_diag}(a) shows the obtained stability regions in the $(k_{1},\, k_{2})$-plane with $l=2.57$ [m] and $d=2$ [m], where the shaded regions are stable regions satisfying \eqref{eqn:stab_cond_nec_suff} with $\kappa_{0}=0$. The dashed orange curve and the dot-dashed yellow curve represent the changes in the stability boundaries for $\kappa_{0}=\overline{\kappa}_{0}/2$ and $\kappa_{0}=\overline{\kappa}_{0}$, respectively.
Furthermore, a sufficient condition independent of road curvatures is practically more useful, which is next provided in the following proposition.

\begin{proposition}\label{prop:stab_cond_suff}
  The desired solution \eqref{eqn:equil_D} of the closed-loop system (\ref{eqn:EOM_gen_point}, \ref{eqn:steer_angle_track}-\ref{eqn:theta0_des}) is stable for any constant $\kappa_{0}$ if either of the following conditions is satisfied:
  \begin{enumerate}
    \item $k_{1}<0$ and $k_{2}>\dfrac{d}{l}\frac{\tan^{2}\gamma_{\max}}{\sqrt{l^{2}+d^{2}\tan^{2}\gamma_{\max}}}$, or
    \vspace{2pt}
    \item $k_{1}>0$ and $k_{2}<-\dfrac{1}{d}$.\vspace{1pt}
  \end{enumerate}
  \IEEEproof See Appendix~\ref{append:proof_prop_1}.
\end{proposition}

\begin{remark}\label{remark:pos_neg_fb}
  The system uses negative feedback if condition 1 is satisfied, and positive feedback if condition 2 is satisfied. For positive feedback, $|k_{2}|$ has to be large to ensure stability when $d$ is small, which may exceed the system actuation capability, and subsequently lead to saturation and oscillations for large errors. This undesired behavior will be shown in the simulations in Section \ref{sec:sim_res}. To ensure the best performance, negative feedback is assumed in the rest of this paper, if not otherwise specified.
\end{remark}

%%%%%%%%%%%%%%% begin figure %%%%%%%%%%%%%%%%%%%
%%%%%%%%%%%%%%% end figure %%%%%%%%%%%%%%%%%%%

To study the tracking performance, we define the transfer function and the corresponding amplification ratio as
\begin{align}
  G(s)&:=\dfrac{E(s)}{K(s)}\,, &
  M(\omega) &:= |G(\textrm{j}\, \omega)|\,,
\end{align}
where $K(s)$ and $E(s)$ are the Laplace transform of $\tilde{\kappa}_{\rm D}$ and $\tilde{e}_{\rm D}$, respectively.
Fourier's theory states that
periodic signals can be represented as an infinite sum of sines and cosines, which can also be extended to absolutely integrable non-periodic signals using Fourier transform. With the assumption that variations on road curvature are absolutely integrable, tracking performance can henceforth be characterized by the amplification ratio from road curvature to lateral deviation at different frequencies according to the superposition principle. From \eqref{eqn:lin_model}, it follows that
\begin{align}
  G(s) &= \mathbf{C}(s\,\mathbf{I}-\mathbf{A})^{-1}\mathbf{B}s \nonumber\\
  &=\dfrac{\frac{V d}{\lambda_{1}^{2}} (1+\frac{d}{l} \lambda_{2} k_{1})s}{s^{2}-\frac{V\lambda_{2} k_{1}}{l\lambda_{1}}(\lambda_{1}+ d k_{2}) s-\frac{V^{2}}{l} \frac{\lambda_{3}}{\lambda_{1}^{2}}}\,,
\end{align}
where
\begin{align}
  \lambda_{4} &=\frac{V^{2}}{l^{2}\lambda_{1}^{2}}\bigr(2 l \lambda_{3} +\lambda_{2}^{2} k_{1}^{2}(\lambda_{1}+ d k_{2})^{2}\bigr)\,.
\end{align}
Thus, the amplification ratio is
\begin{align}
  M(\omega)&=\sqrt{\dfrac{\frac{V^{2} d^{2}}{\lambda_{1}^{4}} (1+\frac{d}{l} \lambda_{2} k_{1})^{2} w^{2}}{w^{4}+\lambda_{4} w^{2}+\frac{V^{4} \lambda_{3}^{2}}{l^{2} \lambda_{1}^{4}}}}\,,\nonumber
\end{align}
which depends on the parameters $(\kappa_{0},\,k_{1},\, k_{2})$.
Fig.~\ref{fig:stab_diag}(b) plots the amplification ratio $M$ for points P, Q and H marked in Fig.~\ref{fig:stab_diag}(a). The green, blue and purple curves represent the results for case P, Q and H, respectively, while the solid, dashed, dot-dashed curves represent the results for $\kappa_{0}=0,\,\overline{\kappa}_{0}/2,\,\overline{\kappa}_{0}$, respectively. The following are observed: (1) the amplification depends more on the gains $(k_{1},\, k_{2})$ than the road curvature $\kappa_{0}$; (2) the amplification ratio of positive feedback is larger than that of negative feedback; (3) the amplification ratio is bounded and minimizing this upper bound can improve the tracking performance of the controller.
One can verify that the magnitude reaches the maximum
\begin{align}\label{eqn:Mag_max}
 M_{\max} %&=\max_{\omega\in \mathbb{R}} M(\omega)\nonumber\\
 &=\dfrac{d}{\lambda_{1}} \cdot
 \dfrac{|l+d \lambda_{2} k_{1}|}{\sqrt{2 l (\lambda_{3} +|\lambda_{3}|) +\lambda_{2}^{2} k_{1}^{2}(\lambda_{1}+ d k_{2})^{2}}}\,,
\end{align}
when
\begin{align}
\omega_{\rm m}&= \dfrac{V}{\lambda_{1}}\sqrt{\dfrac{ | \lambda_{3}|}{l}}\,.
\end{align}
Figs.~\ref{fig:stab_diag}(c, d) show the contours of $M_{\max}$ in the $(k_{1},\, k_{2})$-plane for $d=2,\,3$ [m], respectively, with $l=2.57$ [m] and $\kappa_{0}=0$. One can observe that: (1) The same value of amplification ratio yields narrower region in the case $d>l$ than the case $d<l$, implying that the same gains will have better tracking performance in the latter case; (2) There is a special set of gains highlighted by the black dashed lines that achieves the best tracking performance.

\begin{claim}\label{claim:gain_opt}
    In practice, to achieve reasonably good tracking performance, one can either mount the sensor at the rear axle center point (i.e., $d=0$) or choose $k_{1}=-\frac{l}{d}$ .
    \IEEEproof \eqref{eqn:Mag_max} implies that the amplification ratio from road curvature variations to lateral deviations is always zero at all frequencies when either $d=0$ or
    \begin{align}
     k_{1}&=-\dfrac{l}{d\lambda_{2}} =-\frac{l}{d\big(1+(l^{2}-d^{2})\kappa_{0}^{2}\big)}\,.
    \end{align}
    Based on Claim \ref{claim:lamda2_pos}, road curvature varies in a small neighborhood around $\kappa_{0}=0$. Thus, setting
    \begin{align}
     k_{1}&=-\dfrac{l}{d}\,
    \end{align}
    will give reasonably good tracking performance in practice.
\end{claim}

\section{Simulations \label{sec:sim_res}}
In this section, we perform numerical simulations to demonstrate the effectiveness of the proposed path-following controller. More specifically, we first show the tracking performance under constant curvatures, i.e., the path is either a straight line (${\kappa_0=0}$) or a circle of radius $\rho$ (${\kappa_0 =1/\rho}$). Then we demonstrate that the proposed controller is also capable of following paths with varying curvatures and providing reasonable tracking performance. Meanwhile, we show that the stability conditions obtained in Section~\ref{sec:analysis} can provide insights in the selection of control parameters to achieve desired behaviors. The parameters used in the simulations are provided in Table~\ref{tab:veh_param}. We note that initial conditions with very large lateral deviations are intentionally used to demonstrate the effectiveness on how the controller drives the vehicle to approach the reference path.

%%%%%%%%%%%%%%% begin table   %%%%%%%%%%%%%%%%%%%%%%%%%%
\begin{table}[!ht]
\begin{center}
\rowcolors{1}{LightCyan}{LightMagenta}
\renewcommand{\arraystretch}{1.1}
\begin{tabular}{c||l|c}%{c||l|c|c}
\hline\hline
 \rowcolor{Gray}  & Parameter & Value %& Description
 \\
\hline%\hline
 \cellcolor{white} & $l$ [m]& $2.57$ %&
 \\
 \cellcolor{white}Physical & $d$ [m]& $2 $ %&
 \\
 \cellcolor{white}Parameters &$\gamma_{\max}$ [deg]& $30$ %&
 \\
  \cellcolor{white} & $V$ [m/s]& $20$ %&
 \\
\hline%\hline
 \cellcolor{white}& $k_{1}$ [m/s]& $-0.8$ %&
 \\
 \cellcolor{white}Control& $k_{2}$ [m$^{-1}$]& $0.02$ %&
 \\
 \cellcolor{white}Parameters& $a_{\max}^{\rm lat}$ [m/s$^{2}$] & $4$ %&
 \\
\hline%\hline
 \cellcolor{white} & $\rho$ [m]& $200$ %&
 \\
 \cellcolor{white}Path &$s_{\rm T}$ [m]& $250$ %&
 \\
 \cellcolor{white}Parameters& $N$ & $4$ %&
 \\
 \cellcolor{white}& $\kappa_{\max}$ [m$^{-1}$]& $0.004\pi$ %&
 \\
\hline%\hline
 \cellcolor{white} & $s_{\rm D}(0)$ [m] & $0$ %&
 \\
 \cellcolor{white}Initial & $e_{\rm D}(0)$ [m] & $-10$ %&
 \\
 \cellcolor{white}Conditions & $\theta_{\rm D}(0)$ [deg]& $0$ %&
 \\
\hline\hline
\end{tabular}
\end{center}
\caption{Parameters used in the simulation. The physical parameters are from a Kia Soul 2016 vehicle. \label{tab:veh_param}}
\end{table}
%%%%%%%%%%%%%%%% end table %%%%%%%%%%%%%%%%%%%

\begin{figure}[!t]
  \centering
  \includegraphics[scale=1]{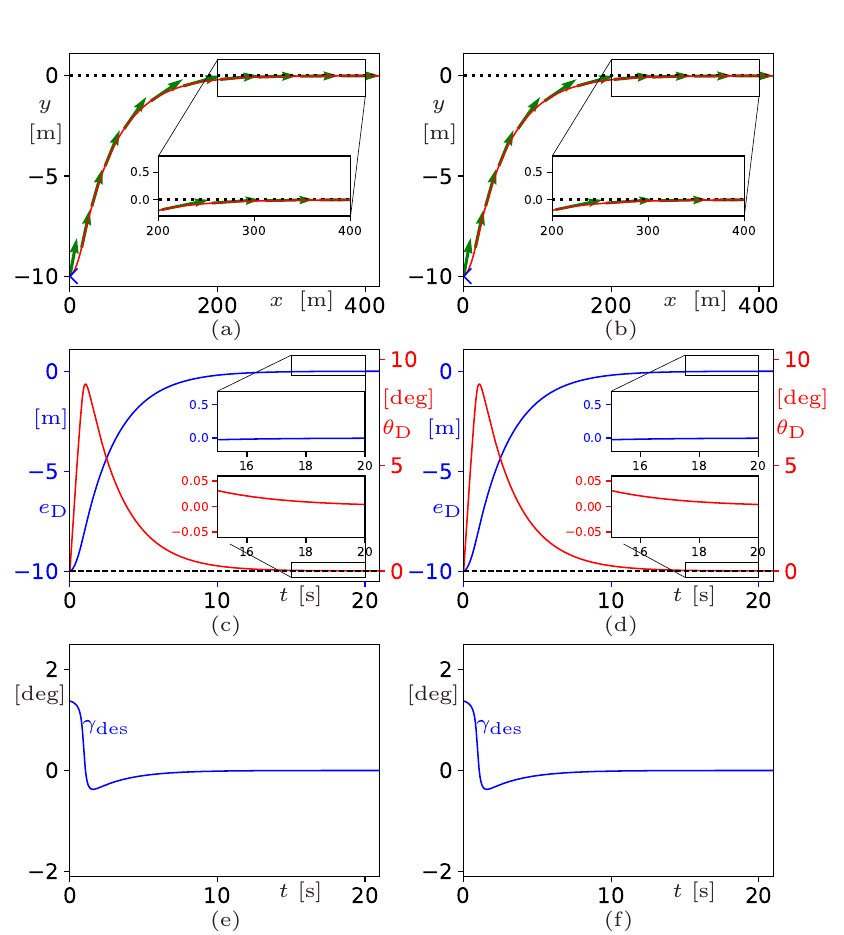}\\
  \caption{Comparison of controller performance to follow a straight path. Left column: sensor location not considered. Right column: sensor location considered. \label{fig:straight_path_comp}}
\end{figure}

\begin{figure}[!t]
  \centering
  \includegraphics[scale=1.0]{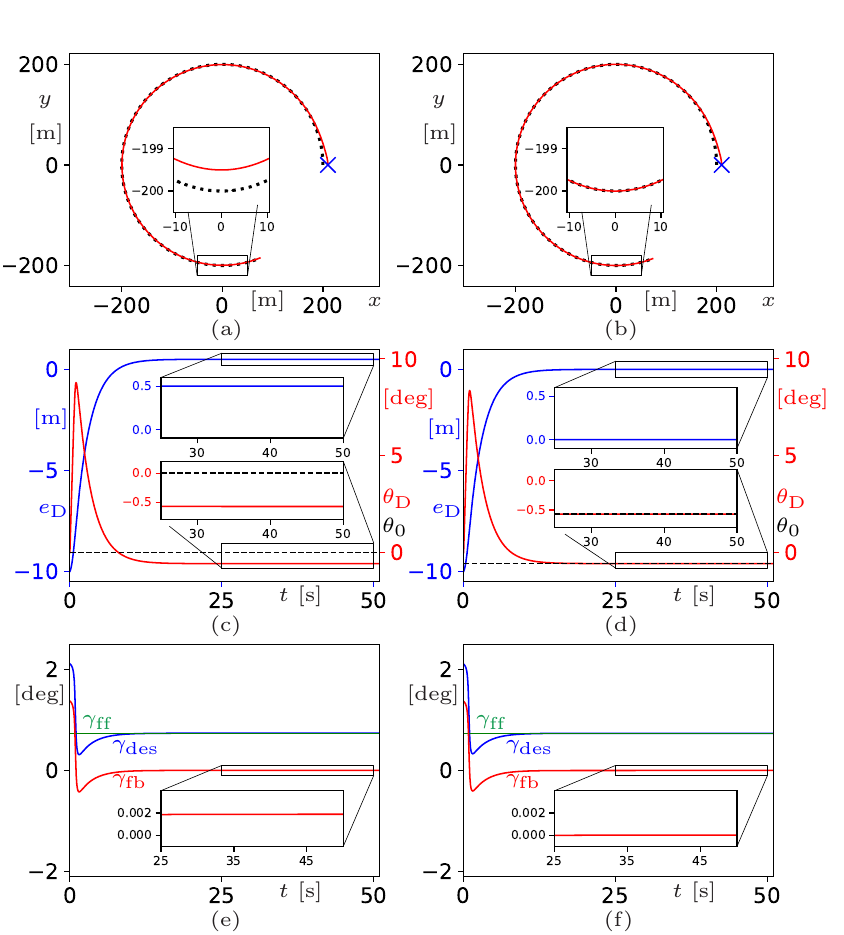}\\
  \caption{Comparison of controller performance to follow a circular path. Left column: sensor location not considered. Right column: sensor location considered. \label{fig:circ_path_comp}}
\end{figure}

\begin{figure}[!t]
  \centering
  \includegraphics[scale=1.0]{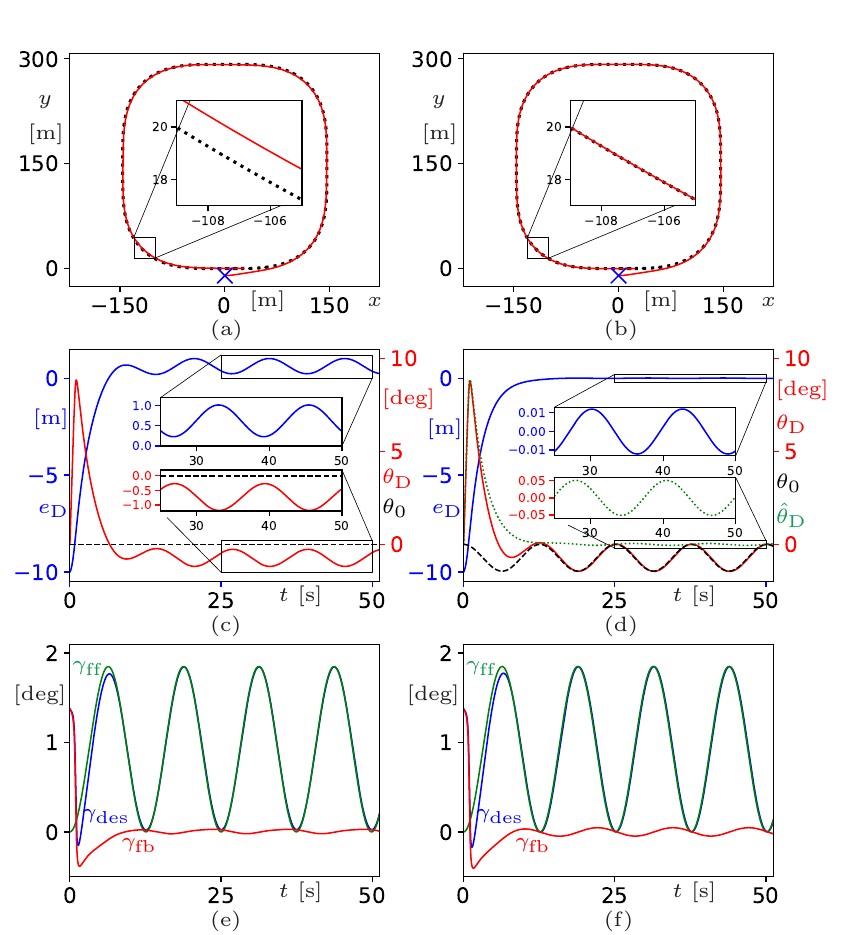}\\
  \caption{Comparison of controller performance to follow a path with varying curvature. Left column: sensor location not considered. Right column: sensor location considered. \label{fig:curv_path_comp}}
\end{figure}

Figs.~\ref{fig:straight_path_comp}, \ref{fig:circ_path_comp} and \ref{fig:curv_path_comp} show the comparison of the tracking performance when the vehicle is trying to follow a straight path, a circular path and a path with varying curvatures, respectively.
In these figures, we maintain the following layout and color scheme.
The left columns correspond to the case that  sensor mounting location is ignored by using controller (\ref{eqn:steer_controller_ff_D_0}, \ref{eqn:steer_controller_fb_D_0}) as the feedforward and feedback law, while the right columns show the case that the sensor location is taken into consideration by using controller (\ref{eqn:steer_controller_ff_sum}, \ref{eqn:steer_controller_fb_sum}). Furthermore, the top panels show the given path and movement of the vehicle (point A) in the $(x, y)$ plane; the middle panels show the time profiles of lateral deviation and yaw angle error; and the bottom panels depict the time profiles of steering angles.
In panels (a) and (b), the dotted black lines indicate the desired path, while the solid red curves represent the position of point A. The green arrows indicate the desired heading given by \eqref{eqn:desired_heading}. In panels (c) and (d), the solid blue curves and red curves represent the lateral deviation $e_{\rm D}$ and yaw angle error $\theta_{\rm D}$, respectively, while the dashed black curves represent the desired yaw angle error $\theta_{0}$, and the dotted green curves represent the difference $\hat{\theta}_{\rm D}$ in \eqref{eqn:theta_hat}. In panels (e) and (f), the solid blue, green and red curves represent the desired steering angle $\gamma_{\rm des}$, the feedforward term $\gamma_{\rm ff}$ and feedback term $\gamma_{\rm fb}$, respectively. Moreover, to highlight the difference in the performance, we use zoomed-in plots overlaid on these panels.

%%%%%%%%%%%%%%% begin figure %%%%%%%%%%%%%%%%%%%
%%%%%%%%%%%%%%% end figure %%%%%%%%%%%%%%%%%%%

Fig.~\ref{fig:straight_path_comp} shows that while following a straight path there is no difference in the performance when sensor location is considered or not, because controllers (\ref{eqn:steer_controller_ff_D_0}, \ref{eqn:steer_controller_fb_D_0}) and (\ref{eqn:steer_controller_ff_sum}, \ref{eqn:steer_controller_fb_sum}) are essentially the same when $\kappa_{\rm D} \equiv 0$.
In Figs.~\ref{fig:straight_path_comp} (a, b), we can observe that the vehicle heading is parallel to the desired heading that is implicated in the feedback term.
Figs.~\ref{fig:straight_path_comp}(c,d) indicate that the vehicle settles down to the given path after a few seconds. Figs.~\ref{fig:straight_path_comp}(e,f) show that the steering reaction is reasonable even for relatively large initial lateral deviation.  Neither overshoot nor oscillations appear as the vehicle approaches the desired path.

%%%%%%%%%%%%%%% begin figure %%%%%%%%%%%%%%%%%%%
%%%%%%%%%%%%%%% end figure %%%%%%%%%%%%%%%%%%%

Fig.~\ref{fig:circ_path_comp} indicates that while following a circular path, the controller (\ref{eqn:steer_controller_ff_D_0}, \ref{eqn:steer_controller_fb_D_0}) always maintains an error in the lateral deviation and yaw angle, while the controller (\ref{eqn:steer_controller_ff_sum}, \ref{eqn:steer_controller_fb_sum}) achieves zero lateral deviation while maintaining desired yaw angle error $\theta_{0}$.
Fig.~\ref{fig:circ_path_comp}(f) shows that the feedback term converges to zero, and the desired steering angle is dominated by the feedforward term that is directly related to the road curvature $1/\rho$. In contrast, when sensor location is ignored, the feedback term converges to a non-zero value (see Fig.~\ref{fig:circ_path_comp}(e)), yielding non-zero lateral deviations and yaw angle errors (see Fig.~\ref{fig:circ_path_comp}(c)).

%%%%%%%%%%%%%%% begin figure %%%%%%%%%%%%%%%%%%%
%%%%%%%%%%%%%%% end figure %%%%%%%%%%%%%%%%%%%

In Fig.~\ref{fig:curv_path_comp}, we validate controllers on the path with varying curvatures, whose curvature varies as a function of the arc length $s$ according to
\begin{align}\label{eqn:curv_s_func}
 \kappa(s) &= \tfrac{\kappa_{\max}}{2}\left(1-\cos\big(\tfrac{2\pi}{s_{\rm T}}s\big)\right),
\end{align}
where $\kappa_{\max}$ is the maximum curvature along the path, and $s_{\rm T}$ is the period in arc length. The path parameters are also provided in Table~\ref{tab:veh_param}. The left column indicates that the controller (\ref{eqn:steer_controller_ff_D_0}, \ref{eqn:steer_controller_fb_D_0}), without incorporating the sensor positions, generates oscillations in the lateral deviation and yaw angle errors when road curvature varies. These oscillations correspond to noticeable vehicular sway motion from side to side while entering or exiting curves, which is commonly misunderstood as improper choice of control gains. In comparison, the right column demonstrates that the controller (\ref{eqn:steer_controller_ff_sum}, \ref{eqn:steer_controller_fb_sum}) achieves satisfactory performance with centimeter accuracy on the lateral deviation. The tiny oscillations caused by the variations of road curvature are generally not noticeable in practice.
Fig.~\ref{fig:curv_path_comp} (f) also shows the following: (1) at the initial stage while approaching the given path, the feedback term $\gamma_{\rm fb}$ is noticeable and dominates the desired steering angle; (2) at the final stage while moving along the path, the feedback term decreases to a small enough value to correct lateral deviations and yaw angle errors, and the feedforward term $\gamma_{\rm ff}$ becomes dominant.

\begin{figure}[!t]
  \centering
  \includegraphics[scale=1.0]{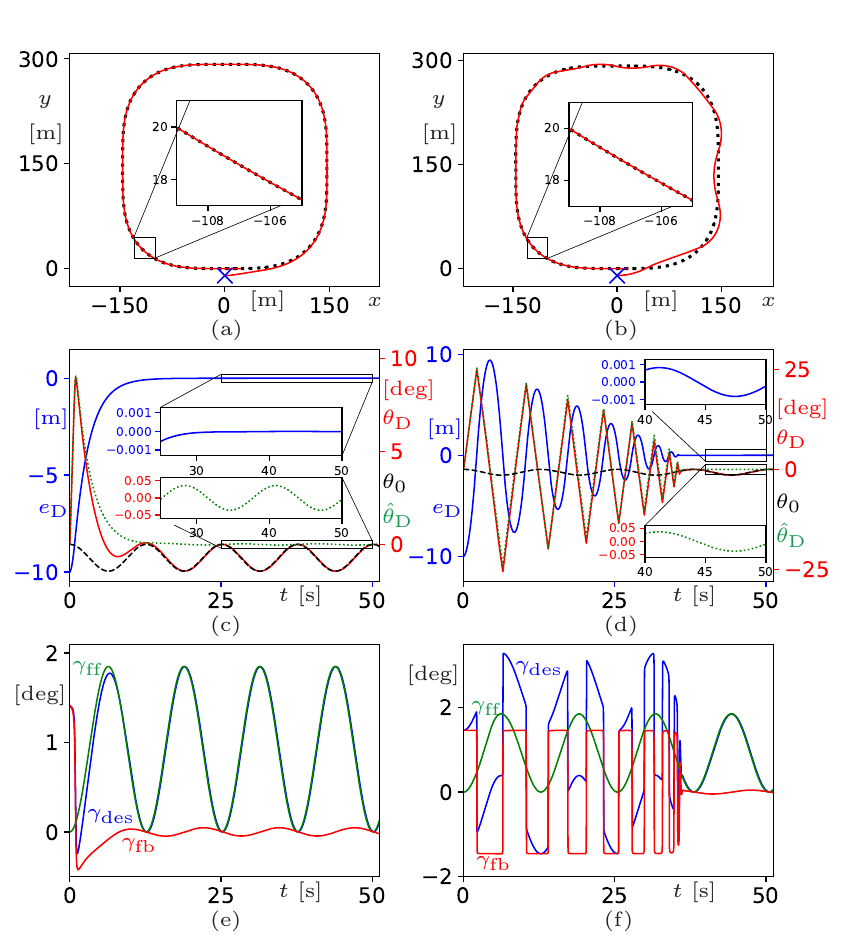}\\
  \caption{Controller performance to follow a path of varying curvature. Left column: negative feedback with $k_{1}=-\frac{l}{d}$ and $k_{2}=0.02\,[\textrm{m}^{-1}]$. Right column: positive feedback with $k_{1}=0.8$ and $k_{2}=-2\,[\textrm{m}^{-1}]$. \label{fig:neg_pos_feedback}}
\end{figure}

In Fig.~\ref{fig:neg_pos_feedback}, we run simulations with the controller (\ref{eqn:steer_controller_ff_sum}, \ref{eqn:steer_controller_fb_sum}) for another two sets of gains when the vehicle is trying to follow the same path as that shown in Fig.~\ref{fig:curv_path_comp}. The left column utilizes the gain $k_{1}$ proposed in Claim~\ref{claim:gain_opt}, while the right column uses positive feedback gains that satisfy the stability condition in Proposition \ref{prop:stab_cond_suff}.  The other parameters remain the same as in Table~\ref{tab:veh_param}, and the color scheme is the same as those for Figs.~\ref{fig:straight_path_comp}, \ref{fig:circ_path_comp}, \ref{fig:curv_path_comp}. The left column indicates that simulation result is consistent with  Claim \ref{claim:gain_opt}, since the lateral deviation converges to zero even when the road curvature varies for this special choice of gains. The right column shows that positive feedback gains satisfying conditions in Proposition \ref{prop:stab_cond_suff} can indeed enable the vehicle to follow the given path when initial errors are small, since Fig.~\ref{fig:neg_pos_feedback}(d) shows reasonably good tracking performance after $t=40$ [s]. However, positive feedback is not robust, and saturations and oscillations will appear when initial errors are large; see Figs.~\ref{fig:neg_pos_feedback}(d,f) when $t<30$ [s]. Therefore, positive feedback should be avoided in practice, which is highlighted in Remark~\ref{remark:pos_neg_fb}.

%%%%%%%%%%%%%%% begin figure %%%%%%%%%%%%%%%%%%%
%%%%%%%%%%%%%%% end figure %%%%%%%%%%%%%%%%%%%

\section{Conclusion}\label{sec:conclusion}

In this paper, we studied the lateral control problem for vehicle path-following and developed a controller to deal with cases where sensors are mounted at arbitrary locations along the vehicle symmetry axis, which is of significant practical importance. We studied stability and tracking performance of the new controller, and provided conditions and guidelines on the selection of control gains. We also used simulations to demonstrate that designing controllers ignoring the sensor mounting positions can generate non-zero lateral deviations for straight/circular paths, as well as unexpected vehicular sway motion for curvy paths. In contrast, we showed that the new controller proposed in this paper can achieve very impressive tracking performance with proper choices of gains. Future research directions may include the integration of steering dynamics, incorporation of longitudinal controllers, consideration of measurement imperfections, validation with experiments on real vehicles, etc.

%%%%%%%%%%%%%%%%%%%%%%%%%%%%%%%%%%%%%%%%%%%%%%%%%%%%%%%%%%%%%%%%%%%%%%%
\section*{ACKNOWLEDGMENT}

The authors would like to thank G\'{a}bor Orosz and D\'enes Tak\'acs for the helpful discussions and insightful suggestions.

\bibliographystyle{IEEEtran}
%\bibliography{Dynamics,LateralControl,Group_pp,Wubing_pp,ITS_Standards} %No space between two file names

\appendix
%\subsection{Proofs \label{append:proofs}}
\subsection{Proof of Claim \ref{claim:lamda2_pos} \label{append:proof_claim_1}}
\begin{itemize}
  \item From Fig.~\ref{fig:circ_path_steer}, we can obtain the maximum curvature that the vehicle is capable of following by maximizing the steering angle, that is
      \begin{align}
        \tan\gamma_{\max} & =\frac{l}{\sqrt{\overline{\kappa}_{0}^{-2} -d^{2}}}\,,
      \end{align}
      yielding $\overline{\kappa}_{0}$ given in the Claim.
%      \begin{align}
%        \overline{\kappa}_{0} & = \dfrac{\tan\gamma_{\max}}{\sqrt{l^{2}+d^{2}\tan^{2}\gamma_{\max}}}\,.
%      \end{align}

  \item The second property is straightforward by noticing that $\lambda_{1}$ in \eqref{eqn:lambda_1_2} is monotonically decreasing with respect to $|\kappa_{0}|$ and $|\kappa_{0}| \le \overline{\kappa}_{0}$.

  \item The third property can be shown by the following two cases.
  \begin{enumerate}
    \item When $l\ge d$, i.e., sensors are mounted behind the front wheel, $\lambda_{2}$ in \eqref{eqn:lambda_1_2} is monotonically increasing with respect to $|\kappa_{0}|$. Thus, $\lambda_{2}\ge 1 >0$.
    \item When $l < d$, i.e., sensors are mounted before the front wheel, $\lambda_{2}$ in \eqref{eqn:lambda_1_2} is monotonically decreasing with respect to $|\kappa_{0}|$. Thus,
        \begin{align}
            \lambda_{2}& \ge 1+(l^{2}-d^{2})\overline{\kappa}_{0}^{2}  \nonumber\\ &=\dfrac{1+\tan^{2}\gamma_{\max}}{1+\frac{d^{2}}{l^{2}}\tan^{2}\gamma_{\max}}>\dfrac{l^{2}}{d^{2}}>0\,.
        \end{align}
  \end{enumerate}
\end{itemize}

\subsection{Proof of Proposition \ref{prop:stab_cond_suff} \label{append:proof_prop_1}}
Condition \eqref{eqn:stab_cond_nec_suff} implies the following two cases.
\begin{itemize}
  \item Case 1:
     \begin{align}\label{eqn:stab_cond_a_1}
      &\left\{
        \begin{array}{ll}
          k_{1} > 0 \,,\\
          k_{2} < -\frac{\lambda_{1}}{d} \,,\\[2mm]
          k_{2} < \frac{d\kappa_{0}^{2}}{\lambda_{1}}+ \frac{l\kappa_{0}^{2}}{\lambda_{1}\lambda_{2}k_{1}}\,.
        \end{array}
      \right.
    \end{align}
    By noticing that
    \begin{align}
      -\frac{1}{d} &\le -\frac{\lambda_{1}}{d} < 0\,, \label{eqn:inequal_case1_1}\\
      0 &< \frac{d\kappa_{0}^{2}}{\lambda_{1}} < \frac{d\kappa_{0}^{2}}{\lambda_{1}}+\frac{l\kappa_{0}^{2}}{\lambda_{1}\lambda_{2}k_{1}}\,,
    \end{align}
    we obtain that \eqref{eqn:stab_cond_a_1} is equivalent to
    \begin{align}\label{eqn:stab_cond_a_2}
        k_{1} &> 0 \,, &
        k_{2} &< -\frac{\lambda_{1}}{d}\,,
    \end{align}
    and its sufficient condition independent of $\kappa_{0}$ is
    \begin{align}\label{eqn:stab_cond_a_3}
      k_{1} &> 0 \,,&
      k_{2} &< -\frac{1}{d}\,.
    \end{align}

  \item Case 2:
  \begin{align}\label{eqn:stab_cond_b_1}
    &\left\{
        \begin{array}{ll}
          k_{1} < 0 \,,\\
          k_{2} > -\frac{\lambda_{1}}{d} \,,\\[2mm]
          k_{2} > \frac{d\kappa_{0}^{2}}{\lambda_{1}}+ \frac{l\kappa_{0}^{2}}{\lambda_{1}\lambda_{2}k_{1}}\,.
        \end{array}
      \right.
    \end{align}
    By noticing that \eqref{eqn:inequal_case1_1} still holds and
    \begin{align}
      \frac{d\kappa_{0}^{2}}{\lambda_{1}} &+ \frac{l\kappa_{0}^{2}}{\lambda_{1}\lambda_{2}k_{1}} < \frac{d\kappa_{0}^{2}}{\lambda_{1}}
      \le \frac{d}{l}\frac{\tan^{2}\gamma_{\max}}{\sqrt{l^{2}+d^{2}\tan^{2}\gamma_{\max}}}\,,
    \end{align}
    one can obtain the sufficient condition for \eqref{eqn:stab_cond_b_1}, that is
    \begin{align}
         k_{1} &< 0 \,,&
         k_{2} &> \frac{d}{l}\frac{\tan^{2}\gamma_{\max}}{\sqrt{l^{2}+d^{2}\tan^{2}\gamma_{\max}}}\,.\label{eqn:stab_cond_b_2}
    \end{align}

\end{itemize}

%%%%%%%%%%%%%%%%%%%%%%%%%%%%%%%%%%%%%%%%%%%%%%%%%%%%%%%%%%%%%%%%%%%%%%%%%%%%%%%%%%%%%%%%%%%%%%%%%%%%
%\begin{IEEEbiography}
%[{\includegraphics[width=1in,height=1.25in,clip,keepaspectratio]{Wubing.jpg}}]{Wubing B. Qin}
%received his BEng degree in School of Mechanical Science and Engineering from
%Huazhong University of Science and Technology, China in 2011, and his MSc degree and PhD degree in
%Mechanical Engineering from the University of Michigan, Ann Arbor in 2016 and 2018, respectively.
%Curretly he is a research engineer at Ford Motor Company. His
%research focuses on dynamics and control of connected automated vehicles, digital systems, ground robotics,
%and nonlinear and stochastic systems with time delays.
%\end{IEEEbiography}
%
%\begin{IEEEbiography}[{\includegraphics[width=1in,height=1.25in,clip,keepaspectratio]{bio_zhaojian}}]{Zhaojian Li}
%Dr. Zhaojian Li is an assistant Professor in the Department of Mechanical Engineering at Michigan State University. He received his Ph.D. from the University of Michigan, Ann Arbor, in 2015. His main research interests include Robotics and Autonomous Vehicles, Intelligent Transportation System, Reinforcement Learning, Vehicle Dynamics, and Optimal Control. He is a senior member of IEEE and recipient of NSF CAREER Award.
%\end{IEEEbiography}
\end{document}